\documentclass[12pt,a4paper]{article}

\usepackage[dvipsnames]{xcolor}
\usepackage{siunitx}
\usepackage{booktabs, multirow, makecell, siunitx}
\usepackage[english]{babel}
\usepackage[authoryear]{natbib}
\usepackage[utf8]{inputenc}
\usepackage{comment}
\usepackage{algorithmic}
\usepackage{algorithm}
\usepackage[official]{eurosym}
\usepackage{multirow}
\usepackage{amsmath, amsthm, amssymb, amsfonts, graphicx,tabularx,rotate}
\usepackage{longtable, lscape, adjustbox,empheq}
\usepackage[figuresright]{rotating}
\usepackage{multicol}
\usepackage{rotfloat,subcaption}
\usepackage[justification=justified, singlelinecheck=false]{caption}
\usepackage{amssymb}
\usepackage{amsmath}
\usepackage{booktabs}
\usepackage{subcaption}
\usepackage{afterpage,url}
\usepackage[tmargin=2.5cm,bmargin=2.25cm,lmargin=2.25cm,rmargin=2.25cm]{geometry}
\usepackage{setspace}

\usepackage[hidelinks]{hyperref}

\usepackage{placeins}

\makeatletter
\AtBeginDocument{%
  \expandafter\renewcommand\expandafter\subsection\expandafter
    {\expandafter\@fb@secFB\subsection}%
  \newcommand\@fb@secFB{\FloatBarrier
    \gdef\@fb@afterHHook{\@fb@topbarrier \gdef\@fb@afterHHook{}}}%
  \g@addto@macro\@afterheading{\@fb@afterHHook}%
  \gdef\@fb@afterHHook{}%
}
\makeatother
\title{\vspace{-60pt} \textbf{Quantifying Demand Shocks in the Green and Digital Transition}}

\author{
Andrea Bastianin\thanks{University of Milan, Italy and Fondazione Eni Enrico Mattei (FEEM). \color{blue}\texttt{andrea.bastianin@unimi.it}}
\and 
Luca Rossini\thanks{University of Milan, Italy and Fondazione Eni Enrico Mattei (FEEM). \color{blue}\texttt{luca.rossini@unimi.it}}
\and 
Marco Zoso\thanks{\textit{Corresponding author}:    University of Milano Bicocca, Italy and Fondazione Eni Enrico Mattei (FEEM). \color{blue}\texttt{m.zoso1@campus.unimib.it}}
}

\date{PRELIMINARY DRAFT \\ \today}

\begin{document}

\maketitle

\doublespacing
\small{\noindent\textbf{Abstract:} We use web search data to construct monthly indexes of derived demand for cobalt, copper, and nickel, which are key inputs in technologies driving the energy and digital transitions. 
We incorporate these indexes into Structural Vector Autoregressive (SVAR) models of global metal markets and identify structural shocks using zero, sign, and magnitude restrictions. 
This approach disentangles supply shocks from several demand-side drivers of metal prices and isolates a transition demand (TD) shock linked to the diffusion of metal-intensive technologies. 
We find that TD shocks generate persistent price effects, especially for copper and nickel, whereas supply and metal-specific demand shocks are more immediate and less persistent. 

\vspace{.5cm}


\vspace{.25cm}

\vfill

\pagenumbering{arabic}
\doublespacing
\normalsize

\clearpage



\clearpage

\section{Introduction}\label{sec:introduction}

Raw materials such as cobalt, copper, and nickel play a central role in the energy and digital transitions, hereafter referred to as the twin transition. 
They are key inputs in technologies such as electric vehicles, renewable energy systems, and data centre infrastructure, and demand for these metals has grown rapidly in recent years. 
At the same time, their supply is geographically concentrated, making their prices increasingly sensitive to shifts in global demand and geopolitical developments. 
Understanding the drivers of metal prices is therefore essential for both macroeconomic analysis and policy.

Data on the production and consumption of most commodities -- including metals, agricultural products, and energy -- are typically released with long publication lags and, in some cases, are only available at an annual frequency. 
This limits their usefulness for real-time monitoring of price drivers and for short- to medium-term analysis of prices and demands \citep{BRTFore}. 
The advent of big data has provided researchers with new data sources and methods to compensate for the lack of timely, high-frequency official statistics \citep{abraham2022big,varian2014big}. 
Leading examples include the use of web search data \citep{choi2012predicting,FerraraSimoni,Moller}, scanner data \citep{Ngscanner}, satellite imagery \citep{donaldson2016view}, and news content \citep{baker2016measuring,barbaglia2023forecasting}.

In this paper, we construct novel monthly Derived Demand Indexes (DDIs) for cobalt, copper, and nickel using web search activity data from Google. 
We extract these indexes from a high-dimensional set of Google Trends series using targeted principal component methods.
Our approach uses search queries related to products that utilize these metals as production inputs, with a particular emphasis on technologies driving the energy and digital transitions. 
Online search volumes provide a measure of consumers’ purchase intentions \citep{goel2010predicting,vosen2011forecasting}. 
For instance, \citet{Moller} show that web search activity captures people’s intentions to buy a house and thereby serves as a proxy for housing demand. 
This interpretation of web search activity is also related to the literature that uses it to proxy investor attention and its impact on asset prices \citep{andrei2015investor,da2011search,vozlyublennaia2014}. 
We therefore build on this literature to interpret our indexes as proxies for derived demand.

The concept of derived demand is based on the idea that the demand for cobalt, copper, and nickel depends on their use as intermediate goods in the production of final goods such as batteries and smart devices \citep{lowinger1984product,verleger2011margin}. 
In our framework the relevant source of final demand may change over time. 
This is consistent with the view that the ``marginal market'' -- that determines the price of a metal -- might change over time as the shares of products in the consumption bundle shift \citep{verleger2011margin}. 
For example, the end-use shares of cobalt have shifted substantially because of cobalt's increased use in battery manufacturing \citep{graedel2022us,miatto2020rise}; therefore, the battery market has become an increasingly important marginal market, whose dynamics affect the price of metals.\footnote{Several papers rely on the notion of derived demand to model energy and metal markets and to forecast their prices \citep[see e.g.][]{baumeister2018product,duarte2021commodity,ederington2019review,issler2014using,stuermer2017industrialization}.}

The resulting DDIs are incorporated into monthly Structural Vector Autoregressive (SVAR) models of the global markets for cobalt, copper, and nickel. 
We identify structural shocks using a combination of zero, sign, and magnitude restrictions, which allows us to disentangle supply shocks from multiple demand-side shocks that drive the real prices of these metals. 
In particular, we isolate a transition demand (TD) shock linked to the uptake of metal-intensive technologies in the energy and digital transitions. 
Our framework enables a quantitative assessment of the relative contribution of each structural driver to price dynamics and highlights the growing macroeconomic relevance of TD shocks.


This paper contributes to the literature developing SVAR models for commodity markets, including those focused on energy \citep{BHaer,KilianAER2009} and metals \citep{romani2024understanding,BaumeisterSpecialFocus}, as well as studies that aim to isolate shocks related to the energy transition \citep{boer2024energy,cross2023drivers}. 
However, to the best of our knowledge, this paper is the first to exploit monthly indexes based on web search data to identify demand shocks related to the energy and digital transitions in SVAR models. 

Our empirical analysis delivers two main findings. 
First, aggregate demand shocks remain a major driver of metal prices, but transition demand (TD) shocks generate persistent price responses that build gradually and decay slowly, especially for copper and nickel.
Second, the importance of TD shocks has increased in recent years, with particular strong evidence for copper and more moderate for nickel.  

The rest of the paper is organized as follows. 
Section \ref{sec:gtrends} describes the construction of DDIs. Section \ref{sec:svar} outlines the identification strategy used in SVAR models. Section~\ref{sec:results} presents the results, and Section \ref{sec:concl} concludes.

\section{Derived demand indexes based on web search data}\label{sec:gtrends}

\subsection{Google Trend Data}\label{sec:DataConstruction}
The absence of reliable data on metal demand, together with the difficulty of tracking such quantities, motivates our construction of a specific index. 
Specifically, we use data from Google Trends (GT) data to construct the monthly derived demand indexes (DDIs) for cobalt, copper, and nickel related to the twin transition.\footnote{Google's share of the global desktop search market has consistently exceeded 85\% over the period 2009--2023, making it the dominant internet search provider \citep{statcounter2026}.} 
GT is an online tool that provides time series of search volumes for specific queries and geographical regions. 
Google does not disclose the total count of queries by keyword; instead, it provides an index that reflects the proportion of searches within a given geographic region and time period. 
The index is then scaled from 0 to 100, with 100 representing the ``peak popularity'' of a search term.

Given that commodity prices are determined at the global level, our analysis considers a set of GT series capturing global search activity sampled at monthly frequency from January 2004 to December 2023. 
We construct the $DDI$ using GT series for products and components that contain battery metals, such as lithium-ion batteries, as well as downstream products that use these batteries, such as smartphones and laptops. 
We consider three main product categories: smart devices (i.e., laptops, smartphones, tablets, and smart vacuums) and related batteries, electric vehicles (EVs), and solar photovoltaic panels (solar PV).

We therefore identify eleven primitive queries related to the purchase or replacement of these products and components: ``buy an electric car'', ``buy solar panels'', ``buy lithium battery'', ``buy a smartphone'', ``buy a laptop'', ``buy a tablet'', ``replace phone battery'', ``replace laptop battery'', ``replace tablet battery'', ``buy a powerbank'', and ``buy robot vacuums''. 
These queries capture consumer demand for products related to the twin transition containing cobalt, copper, and nickel. 
When a keyword is entered into GT, the platform suggests up to 25 semantically related queries.
We therefore expand our initial list by adding related queries for each primitive query. 
After manually removing unrelated queries, the resulting set contains 121 search terms. 
This set is further reduced by dropping low-volume queries, defined as GT series with more than 5\% of their values equal to zero. 
This yields a total of 108 GT series, which are grouped into three main categories capturing the search activity for smart devices and related batteries (84 GT), EVs (9 GT), and solar PV (15 GT).

GT data are based on a random sample of Google search activity and are not perfectly reproducible over time. 
GT series change over time and vary depending on the Internet Protocol address used to extract the data. 
To address this issue, we retrieved each GT series twelve times, using three machines and collecting the data on different days, and computed the sample average across these samples \citep[see][]{Eichnauer,Medeiros}.

\subsection{Construction of the Derived Demand Indexes}

Once the GT series are collected, we use Principal Component Analysis (PCA) methods to summarise the information contained in GT series into a single metal-specific factor, which we interpret as our DDIs. 
Because PCA methods are not sensitive to non-stationarity, the GT series must first be preprocessed 
\citep[see e.g.][]{Borup,choi2012predicting,Marinucci,FerraraSimoni,Moller,Yu}.

We initially address the large volatility in the GT series by log-transforming each series as $gt_{i,t} = \ln\left(1+GT_{i,t}\right)$ and then apply the  \citet{hamilton2018you} filter to extract the cyclical component of $gt_{i,t}$, where $i$ denotes individual GT series. %
%
As shown by \citet{hamilton2018you}, this approach can be applied to both seasonally and non-seasonally adjusted data and is robust to the seasonal adjustment procedure. 
The method therefore allows us to remove seasonality and induce stationarity before PCA without taking a stance on the nature of trends or the form of seasonality characterizing GT series.\footnote{We also considered the pre-processing approach in \citet{Borup} and \citet{Moller}. This is based on a sequential testing method to distinguish between stationarity, deterministic trend and unit root behavior; however, it is not effective in our setting, as some of the GT still showed a trend after the procedure was applied.} 

We then construct the DDIs using Targeted PCA (TPCA). 
This method, introduced by \citet{BaiNg}, aims to filter out the noise from the GT data by reducing the number of variables fed into the PCA, thereby improving the estimation of the latent factor. 
TPCA proceeds in two steps: (\textit{i}) the selection of GT series targeted to the variable of interest, (\textit{ii}) the computation of a latent factor via standard PCA. 
In the first step, we select relevant GT series with a soft-thresholding rule based on the Elastic-Net of \citet{zou2005regularization}.

Let $\widetilde{\mathbf{gt}}_t = \left(\hat{\nu}_{1,t},\hat{\nu}_{2,t},\ldots,\hat{\nu}_{G,t}\right)^\prime$ be the $(G\times1)$ vector containing the cyclical component of the GT series extracted with the method of \citet{hamilton2018you}, with $G=108$. 
We use the Elastic Net to select a subset of query series $\widetilde{\mathbf{gt}}_{m,t}\in \widetilde{\mathbf{gt}}_t$ that are most relevant for explaining the cyclical component of the log real price of metal $m$, extracted using the Hamilton filter. 
%
%
Following \citet{Borup} and \citet{Moller}, we tune the Elastic Net parameters to assign equal weight to the LASSO and ridge penalty ($\alpha=0.5$), and to return a vector of ten GT series for each metal.\footnote{We select ten series as a standard choice in the Google Trend literature. In the Supplement (available from the authors upon request), we show that the results are robust when selecting the number of series with an information criterion.} 
This procedure yields three possibly overlapping sets of selected queries: $\widetilde{\mathbf{gt}}_{m,t}$ for $m=\text{cobalt (CO), copper (CU), and nickel (NI)}$. The metal-specific DDIs are then estimated as the first principal component of each of these vectors and are denoted as $DDI_{CO,t},DDI_{CU,t}$, and $DDI_{NI,t}$.

\subsection{Validation of the DDI}

Before using the DDIs in the SVAR analysis, we evaluate whether they capture transition-related demand for metals.  
We do so by comparing the indexes with metal prices and, in the Supplement, with sales data and available measures of metal consumption.

\begin{figure}[h!]
    \centering
    \caption{Correlation between DDI and the cyclical component of real prices from 12/2006 to 12/2023.}    
    \includegraphics[width=0.4\textwidth]{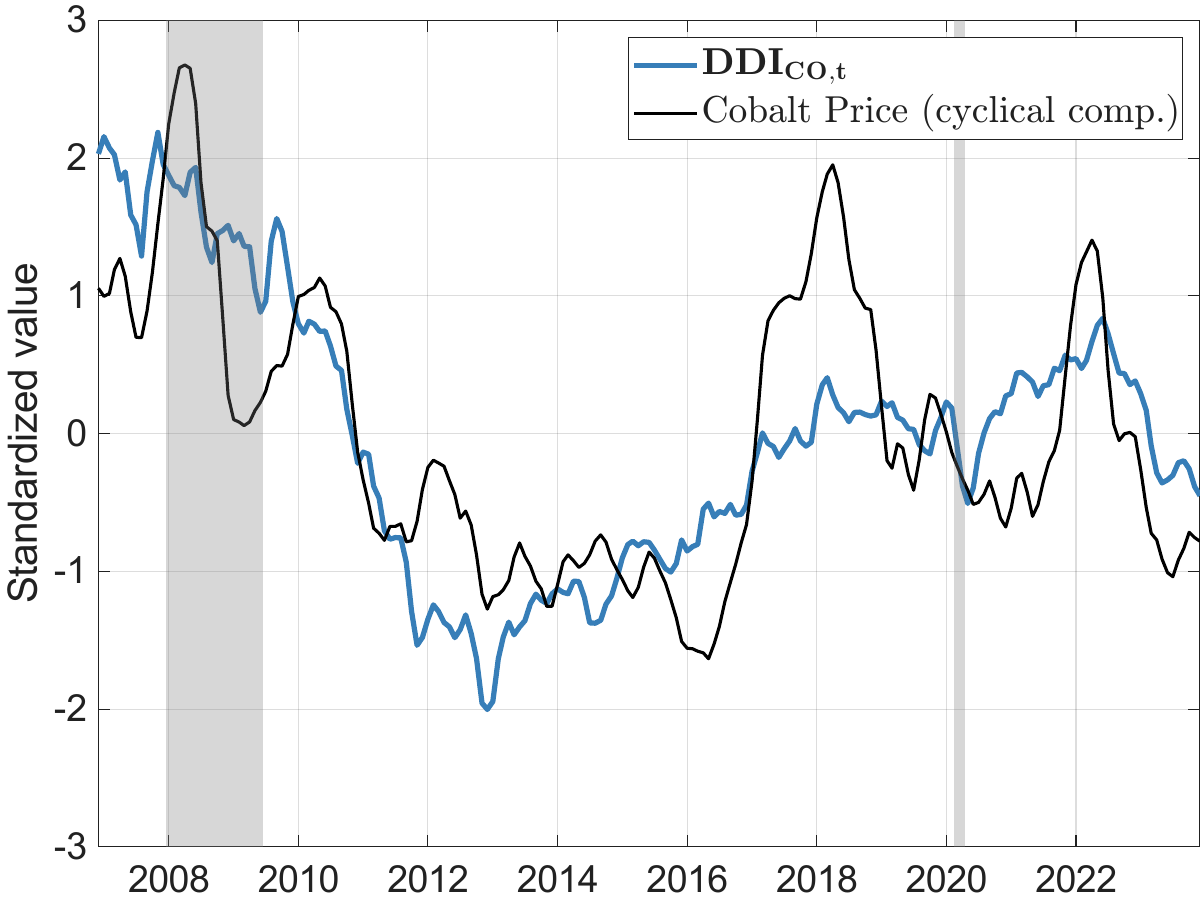}
    \includegraphics[width=0.4\textwidth]{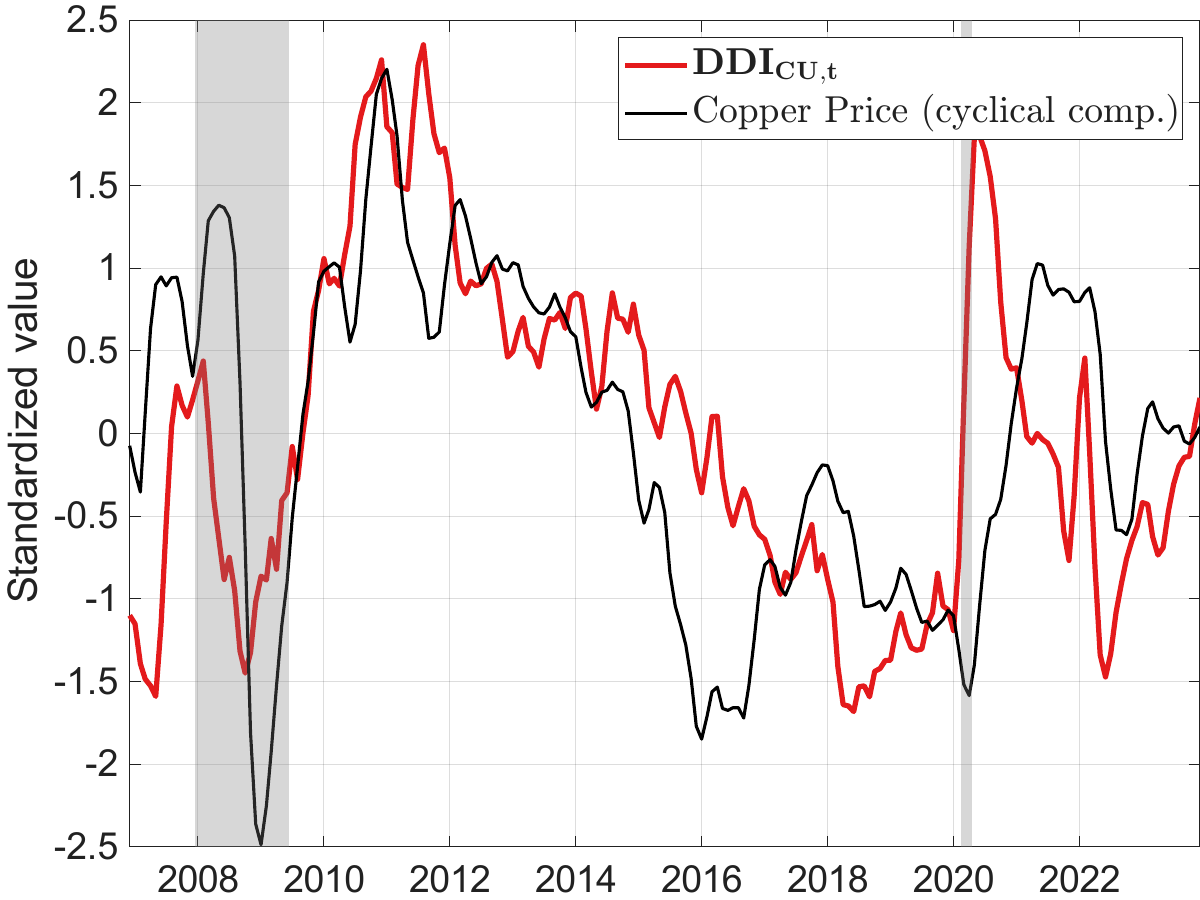}\\
    \includegraphics[width=0.4\textwidth]{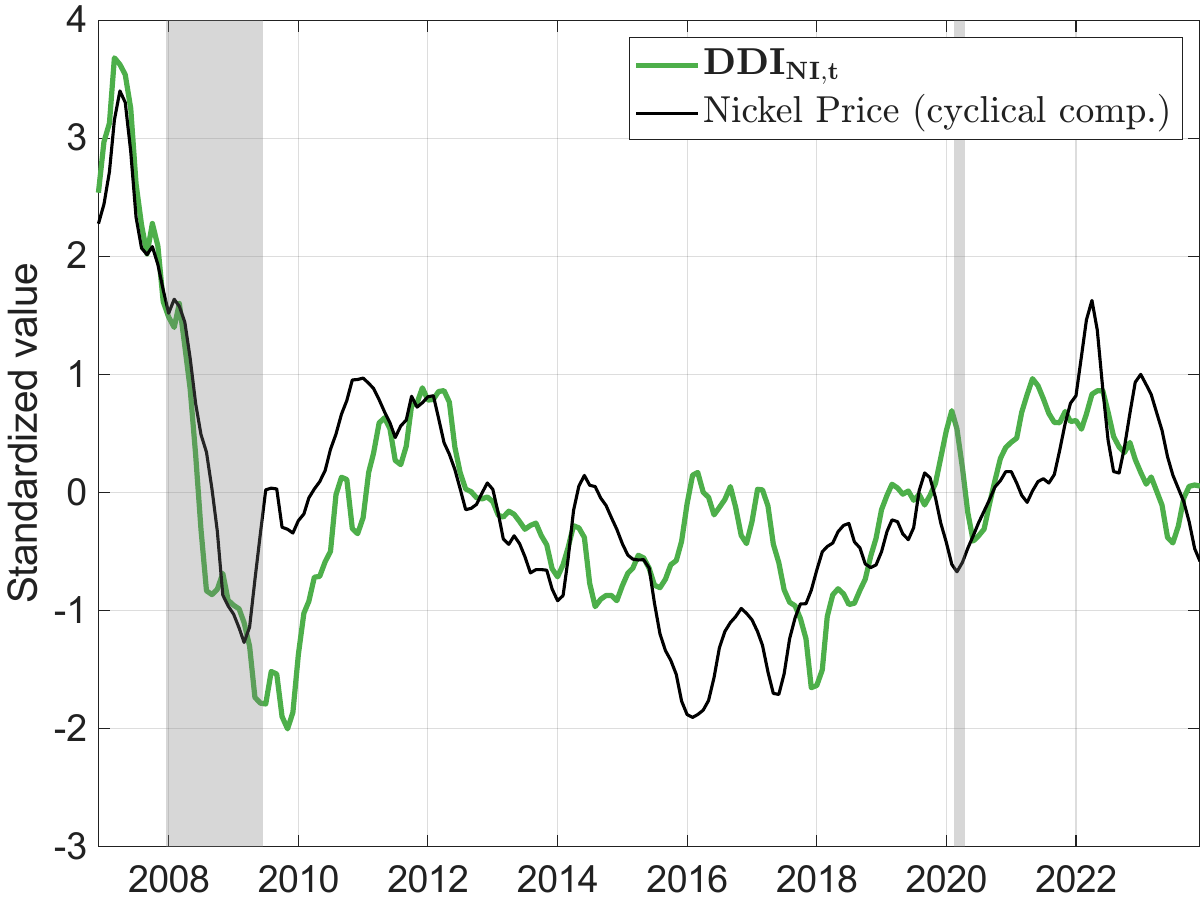}
    \label{fig:FigCorrPrices} 
    \caption*{\scriptsize\textit{Notes:} all series are standardized to have zero mean and unit standard deviation. The plotted lines represent three-period moving averages to smooth short-term fluctuations. Grey shaded areas indicate National Bureau of Economic Research (NBER) recessions.} 
\end{figure}

Figure \ref{fig:FigCorrPrices} plots, for each metal, the DDI (colored line) together with the cyclical component of the corresponding log real price (black line), where the latter is extracted using the \citet{hamilton2018you} filter. 
In all cases, the two series exhibit a strong contemporaneous correlation, supporting the information content of the DDIs and the effectiveness of the TPCA method.
Specifically, cobalt shows the strongest correlation (74.5\%), followed by nickel (72.1\%), while the correlation for copper is 49.6\%. 
The top left panel illustrates how $DDI_{CO,t}$ reflects the sharp price surges in 2022, driven by soaring demand for metals used in EV production and a sluggish supply response. 
Similarly, the bottom panel shows that $DDI_{NI,t}$ captures the trough in nickel prices in 2008, associated with the global financial crisis. 

In all panels, a clear discontinuity emerges during the COVID-19 pandemic. 
This is confirmed by the fact that excluding the COVID-19 period increases the correlation between the DDIs and prices for all three metals. 
The effect is particularly pronounced for copper, for which the correlation rises by about 13\%. 
This pattern suggests that the relationship between search-based indicators and the underlying economic mechanisms temporarily weakened.
During the pandemic, mobility restrictions and stay-at-home policies increased online search activity, thereby raising DDIs even as underlying metal demand weakened with the contraction in global economic activity. 
At the same time, metal prices declined because industrial demand collapsed, while inventories likely buffered short-run supply adjustments. 
Consequently, the usual co-movement between DDIs and metal prices temporarily weakened because search activity and realized demand moved in opposite directions. Additional validation results are reported in the Supplement (available from the authors upon request).

\FloatBarrier

\section{Structural VAR models for global metal markets}\label{sec:svar}

We incorporate the DDIs into metal-specific SVAR models to identify supply- and demand-side drivers of the real price of each commodity:

\begin{equation}
    \label{eq:svar}
    \mathbf{A}_{m,0} \mathbf{y}_{m,t} = \mathbf{c}_m + \sum_{j=1}^p \mathbf{A}_{m,j} 
    \mathbf{y}_{m,t-j} + \boldsymbol{\varepsilon}_{m,t} \qquad \boldsymbol{\varepsilon}_{m,t} \sim \mathcal{N}(\mathbf{0},\mathbf{I}_n) \ \text{for} \ m \in \{\text{CO},\text{CU},\text{NI}\},
\end{equation}
where $n$ denotes the number of endogenous variables, $\mathbf{A}_{m,0}$ captures the contemporaneous relations among the model variables, while $\mathbf{c}_m$ and $\mathbf{A}_{m,j}$ are the vector of intercepts and the matrix of coefficients, respectively.

The vector of endogenous variables is defined as $\mathbf{y}_{m,t} = \left[\Delta prod_{m,t}^o, \Delta WIP_t,p_{m,t},DDI_{m,t} \right]^\prime$ for $m \in \{\text{CO},\text{CU},\text{NI}\}$, where $\Delta prod_{m,t}^o$ denotes the percentage growth rate of global metal ore production\footnote{As robustness checks, the production of refined metals is also considered. For instance, focusing on nickel, the distinction between ore and refined production reflects the difference between raw material extraction and processed output, with ore data capturing gross volumes of mined material and refined data representing the nickel content available for industrial use, thereby offering a more accurate measure of supply relevant to downstream economic activity.}, which is sourced from the WBMS Database. 
For cobalt, we use the sum of ore production for nickel and copper given that 99\% of cobalt is mined as a byproduct of these metals \citep{cobaltinstitute2024cobalt}. 
$\Delta WIP_t$ is the percentage growth rate of World Industrial Production \citep{BHaer}. 
$p_{m,t}$ is the natural logarithm of the global real price of commodity $m$, obtained by deflating the nominal price using the US CPI from the Federal Reserve Economic Data (FRED), and expressed in constant 2023 US dollars. 
Specifically, for cobalt we use the London Metal Exchange (LME) spot price, whereas for copper and nickel we use the LME spot price adjusted for delivery to European ports (CIF). 
All price series are sourced from the International Monetary Fund (IMF) website. 
Finally, $DDI_{m,t}$ denotes the derived demand index introduced in Section~\ref{sec:gtrends}.

In the empirical analysis, we estimate all models with monthly data over the period from December 2006 to December 2023 yielding 205 observations. 
The VAR lag order is set to $p=3$ (other robustness checks using $p=6$ and $p=12$ yield qualitatively similar structural impulse responses). 
All the models are estimated using the Bayesian methodology detailed in \citet{AriasRubioWaggoner2018} and imposing a diffuse Normal–Inverse-Wishart prior for the reduced-form parameters. 
For further details about the estimation procedure we refer the reader to the Supplement (available from the authors upon request).

\subsection{Identification}\label{sec:identification1}

Starting from Equation~\eqref{eq:svar}, we identify four structural shocks by combining:
(\textit{i}) a set of contemporaneous sign and zero restrictions imposed on the matrix of impact responses $\mathbf{A}_{m,0}^{-1}$; and
(\textit{ii}) restrictions on the magnitude of the response of real economic activity to what we label transition and aggregate demand shocks.

\bigskip 

\noindent{\textit{Contemporaneous zero and sign restrictions.}} 
As shown in Table \ref{tab:zerosignrestr}, we identify four structural shocks by enforcing a set of contemporaneous sign and zero restrictions. 
We normalize shocks so that they are all associated with an increase in the real price of metals (i.e., a negative supply and positive demand shocks).

\begin{table}[h!]
    \centering
    \caption{Sign and zero restrictions on the impact response matrix $\mathbf{A}_{m,0}^{-1}$}
\begin{tabular}{c|cccc}
    \hline
           & \multicolumn{4}{c}{shock} \\\cline{2-5}
           & (1) & (2) & (3) & (4) \\
          variable & Supply & Aggregate Demand & Transition Demand & Metal-specific Demand \\\hline
         $\Delta prod_{m,t}^o$ & - & + & $\ast$ & + \\
         $\Delta WIP_{t}$ & - & + & $\ast$ & - \\
         $p_{m,t}$ & + & + & + & + \\
         $DDI_{m,t}$ & 0 & $\ast$ & + & 0 \\\hline
\end{tabular}
    \label{tab:zerosignrestr}
        \caption*{\scriptsize\textit{Notes}: ``$\ast$'' indicates that the element is left unrestricted. The signs follow the convention that all shocks are normalized to generate an increase in the real price of metals (i.e., a negative supply shock and positive demand shocks).}
\end{table}

\begin{itemize}
    \item \textit{Supply shock}. A negative supply shock represents an exogenous disruption to the production process -- such as labor strikes or natural disasters affecting major mining operations -- that leads to a contraction in global metal production within the same month. 
    This reduction in supply exerts upward pressure on real metal prices through reduced availability, as commonly assumed in the literature \citep{BRT26Copper,BaumeisterSpecialFocus,BamueisterPeersman2013,KilianMurphy2012}. 
    Importantly, we assume that the within-month response of $DDI_{m,t}$ is constrained to be zero. This assumption is motivated by two considerations: either by appealing to the notion that inventories smooth short-term fluctuations in final goods demand, or by considering that consumers’ intention to purchase products included in the DDI is unaffected on impact.
    
    \item \textit{Aggregate demand} (AD) \textit{shock}. An AD shock is associated with fluctuations in the global business cycle and the resulting changes on the demand for industrial commodities. 
    We follow the common assumption that an unanticipated aggregate demand expansion leads to an increase in global metal production, global economic activity, and the real price of metals \citep[see e.g.][]{BaumeisterSpecialFocus,BHaer,boer2024energy,KilianMurphy2012}. 
    In this case, the impact response of $DDI_{m,t}$ is left unrestricted.
    
    \item \textit{Transition demand} (TD) \textit{shock}. This shock captures unexpected increases in demand for metals driven by their role as inputs in selected final goods associated with the energy and digital transition -- such as batteries, electric vehicles, solar panels, and smart devices. 
    Conceptually, it represents a metal-specific demand shock arising from surges in demand for these goods.
    We assume that a positive TD shock leads to an immediate increase in real metal prices and $DDI_{m,t}$, while remaining agnostic about its contemporaneous effects on production and economic activity.
    
    \item \textit{Metal-specific demand} (MD) \textit{shock}. A MD shock captures unexpected changes in the demand for a specific metal that are not directly linked to the global business cycle or to the energy and digital transition. 
    These shocks are a residual category encompassing innovations, such as precautionary demand shocks (e.g., anticipation of future supply disruptions), expectation shocks related to shifts in uncertainty, and unforeseen increases in demand arising from uses of the metal not captured by our DDIs. 
    These shocks are assumed to have no contemporaneous impact on $DDI_{m,t}$, while generating an immediate increase in real metal prices and global production, and simultaneously inducing a contraction in global economic activity due to the cost-push effects of higher input prices \citep[see e.g,][]{boer2024energy,BaumeisterSpecialFocus,KilianMurphy2012}.
\end{itemize}

\bigskip

\noindent{\textit{Dynamic ``magnitude'' restrictions.}} Table~\ref{tab:zerosignrestr} shows that the identifying restrictions  do not rule out cases in which the impact responses associated with AD and TD shocks share the same sign (see Column 2 and 3). 
In such situations, the identification of AD and TD shocks would fail. 
To avoid any potential confounding between AD and TD shocks, we impose an additional restriction: the response of real economic activity to an AD shock must be larger in magnitude than the response to a TD shock, both on impact and during the following three months.
The idea of restricting the magnitude of a variable’s response to different shocks is also employed by \citet{cross2022role} to disentangle precautionary and speculative oil demand shocks.\footnote{Related approaches include \citet{piffer}, who propose restricting the shape of structural impulse responses, and \citet{amir2021identification}, who develop a framework for enforcing ranking restrictions.}

\section{Results}\label{sec:results}



In this section, we report impulse response function, forecast error variance decomposition, and historical decomposition for the shocks identified in Section~\ref{sec:identification1}.

\subsection{Structural impulse response analysis}\label{sec:irf1}

We compute structural impulse-response functions (IRFs)  from 2,500 posterior draws that satisfy the identification restrictions described in Section \ref{sec:identification1}. 
We report impulse responses over a 24-month horizon to a one-standard-deviation shock and normalize the sign so that all shocks generate a positive price response (i.e., we consider a negative supply shock and positive demand shocks). 
We follow the Bayesian decision-theoretic approach of \citet{INOUE2022457}, which conducts inference on the entire vector of impulse responses and therefore accounts for dependence across horizons and variables.\footnote{This differs from the conventional approach based on pointwise posterior medians and marginal credible intervals, which ignore this dependence and may imply response paths that are inconsistent with any feasible parameterization of the underlying VAR model. Our implementation uses an angular loss function, which emphasizes similarity in dynamic response paths and is invariant to the scaling of both the data and the normalization of structural shocks.} 
The Bayes estimator corresponds to the impulse-response vector that minimizes expected loss under the joint posterior distribution. 
Estimation uncertainty is summarized using joint credible sets defined as the subset of posterior draws with the lowest posterior risk whose cumulative probability equals the desired credibility level. 
Figures \ref{fig:irfs_cobalt}--\ref{fig:irfs_nickel} present, for each metal, the impulse responses of ore production, WIP, and the real price to the four structural shocks. IRFs for variables entering the model in first differences are cumulated. 
Each panel reports the Bayes estimate (thick blue line), the posterior trajectories in the 68\% joint credible set (thin gray lines), and, for comparison, conventional 68\% pointwise credible intervals (dashed lines). 

\bigskip

\noindent\textit{Supply and AD shocks.} A natural starting point is the response to supply and aggregate demand (AD) shocks, whose properties are well-established in the commodity market literature. 
AD shocks generate the largest impact responses and remain among the most persistent shocks, with impact price responses (based on the Bayes estimate) of 6.50\% for cobalt, 3.67\% for copper, and 7.05\% for nickel, reaching peak responses of 8.29\%, 4.67\%, and 9.29\% within the first 4 months, in all cases. 
Consistent with their business-cycle interpretation, AD shocks generate a simultaneous expansion in both ore production and World IP. 
Supply shocks generate the second largest impact responses, with price responses of 5.54\% for cobalt, 3.36\% for copper, and 2.55\% for nickel, accompanied by production contractions and negative World IP responses across all metals. 
Both patterns are consistent with results for the oil market \citep{BHaer} and for metals \citep{bauer2024critical,boer2024energy}.

\bigskip

\begin{figure}[t]
    \centering
        \caption{Impulse responses of cobalt production, world industrial production and real cobalt price to the supply, AD, MD, and TD shocks.}
    \includegraphics[width=0.8\textwidth]{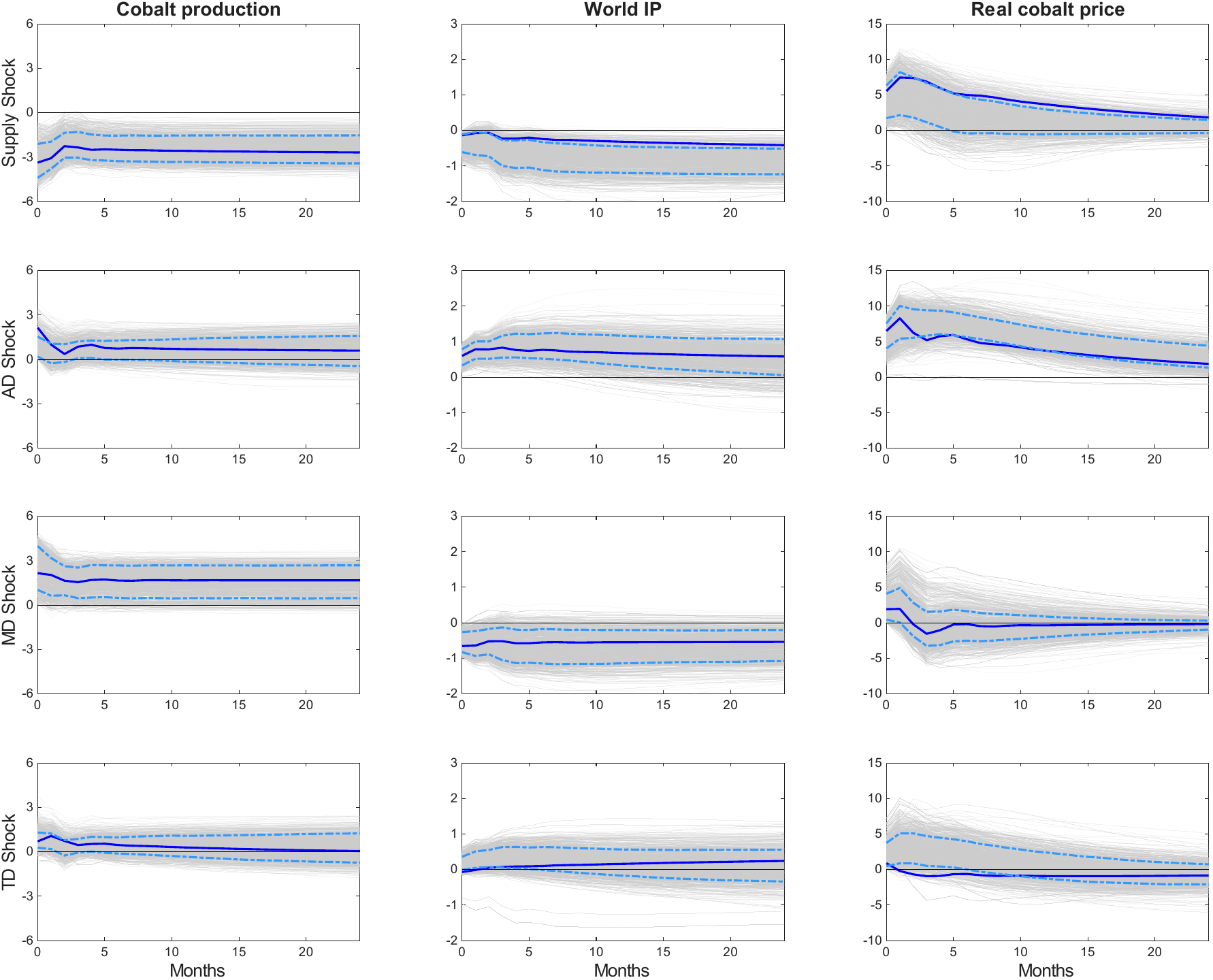}
    \label{fig:irfs_cobalt}
    \caption*{\scriptsize \textit{Notes}: The plots are based on 2500 draws showing the 68\% pointwise credible sets (light blue dashed lines), the Bayes estimator under angular loss function (blue lines) and the 68\% joint credible set under the same loss function (gray lines). The full set of impulse responses is available in the Supplement (available from the authors upon request). The size of all shocks is one standard deviation. The response of variables entering the model in first differences has been accumulated to get log-levels.}
\end{figure}

\begin{figure}[t]
    \centering
        \caption{Impulse responses of copper production, world industrial production and real copper price to the 4 shocks.}
    \includegraphics[width=0.8\textwidth]{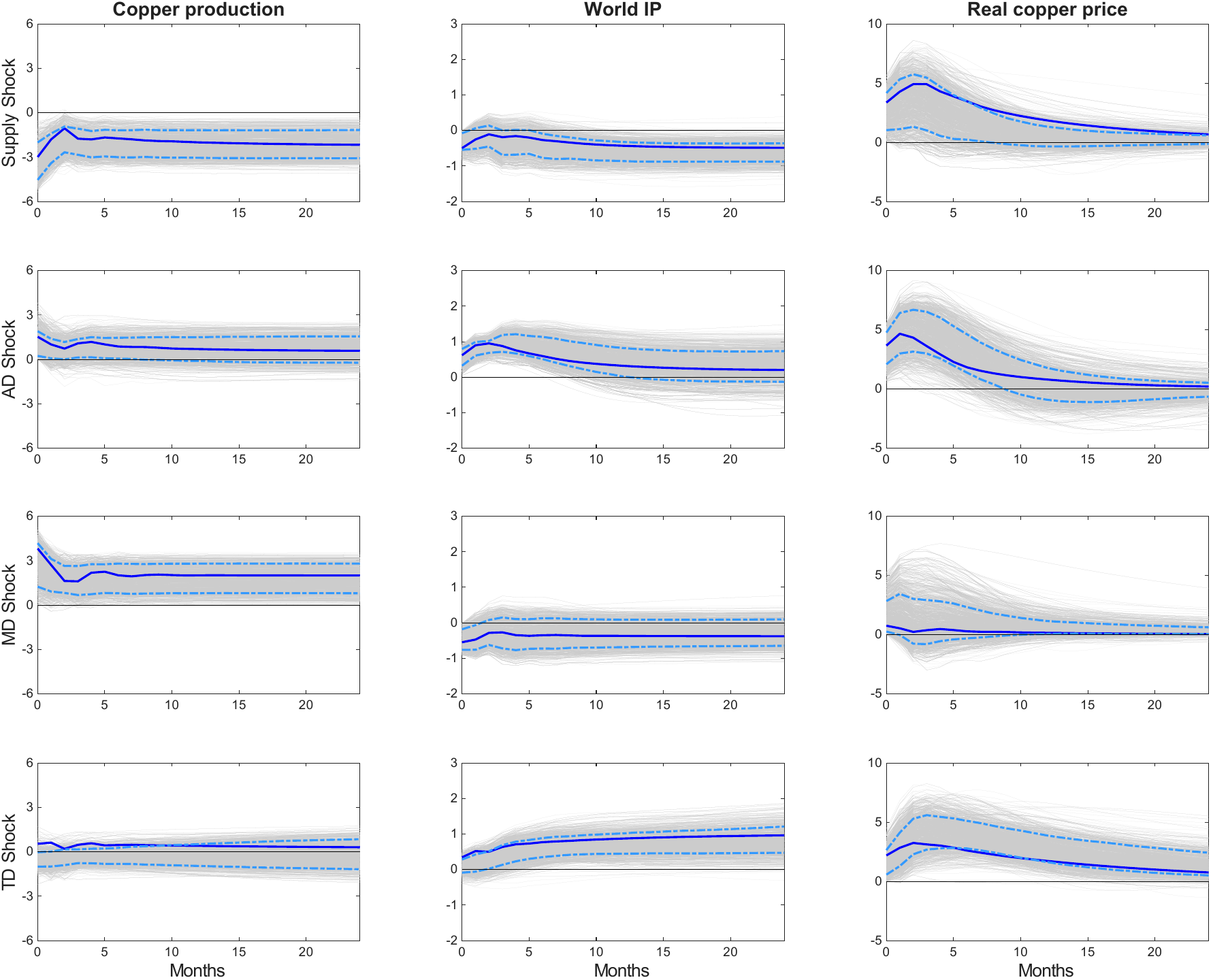}
    \label{fig:irfs_copper}
    \caption*{\scriptsize \textit{Notes}: See notes to Figure \ref{fig:irfs_cobalt}.}
\end{figure}

\begin{figure}[t]
    \centering
        \caption{Impulse responses of nickel production, world industrial production and real nickel price to the 4 shocks.}    \includegraphics[width=0.8\textwidth]{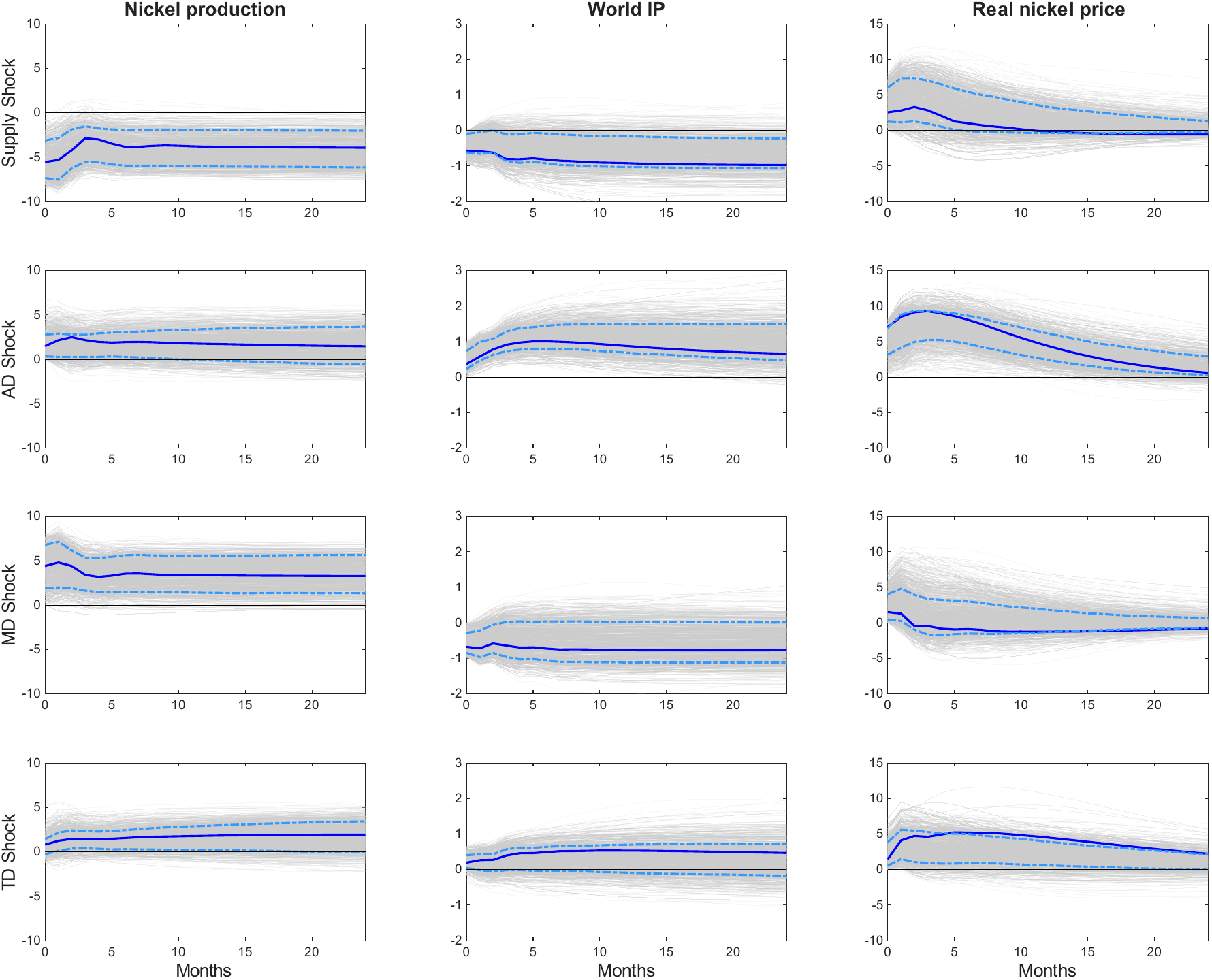}
    \label{fig:irfs_nickel}
    \caption*{\scriptsize \textit{Notes}: See notes to Figure \ref{fig:irfs_cobalt}.}
\end{figure}

\noindent\textit{TD shocks.} TD shocks exhibit a pattern that differs from the other shocks in the model, and that is directly informative about the macroeconomic nature of the twin transition. 
Specifically, TD shocks have two distinctive features. 
First, they generate persistent and economically significant price effects that build up over time. 
On impact, price responses are moderate: 0.89\% for cobalt, 2.19\% for copper, and 1.45\% for nickel.
However, the response to TD shocks strengthens after impact: peak responses are substantially larger than the corresponding impact responses and are reached only after several months. In particular, the peaks are at 3.24\% for copper at month 3 and 5.23\% for nickel at month 6 later than the peaks associated with the other shocks.
The ratio of peak to impact response is particularly striking for nickel, at 3.6, compared to 1.3 for AD shocks and 1.3 for supply shocks. 
For copper and nickel, TD impulse responses decline more slowly than those of supply shocks, despite their smaller initial effect.
A different story arises for cobalt, where the evidence for a sustained price build-up following TD shocks is considerably weaker than for copper and nickel.

Second, and in sharp contrast to their price effects, TD shocks have mostly little effect on metal production or on WIP. 
Impact responses for ore production are 0.70\%, 0.54\%, and 0.80\% for cobalt, copper, and nickel -- well below those of supply and AD shocks -- while WIP responses of -0.07\%, 0.34\%, and 0.19\% are negligible and their credible intervals always include zero. 
These findings are consistent with the interpretation of TD shocks. 
Demand generated by the energy and digital transition for specific metal-intensive final goods -- batteries, electric vehicles, solar panels -- that is large enough to move metal prices persistently but does not constitute a sufficient share of global output to shift aggregate real activity on impact.

\bigskip

\noindent\textit{MD shocks.} MD shocks are characterized by an absence of price amplification beyond impact: peak price responses occur at month 1 for all three metals, implying that the entire effect materializes on impact and then quickly disappears. 
Impact price responses are smaller than those of AD and supply shocks and, for copper and nickel, smaller than the peak TD responses documented above. 
This pattern is particularly notable given that MD shocks are associated with large production responses, combined with WIP contractions, consistent with a cost-push channel. 
The absence of persistent price dynamics distinguishes MD shocks sharply from TD shocks, which build up over several months, and confirms that the decomposition captures economically distinct phenomena rather than observationally equivalent demand shocks.

\bigskip

\noindent\textit{Cross-metal heterogeneity.} Two patterns of cross-metal heterogeneity are worth highlighting. 
The first concerns cobalt. 
Unlike copper and nickel, the cobalt TD shock generates  a flat and uncertain response: the Bayes estimate peaks on impact and the joint credible set remains wide at all horizons. 
One possible explanation is cobalt's byproduct nature -- since no mine is dedicated exclusively to cobalt extraction, its production is tied to that of nickel and copper, which may make supply less responsive to metal-specific demand shifts and amplify the role of supply-side noise in the price signal. 
The second concerns the speed of price adjustment under TD shocks. 
The peak response for nickel occurs at month 6, twice as late as for copper (month 3), and the price profile remains elevated throughout the 24-month horizon. 
Given nickel's central role in battery cathodes for electric vehicles, this slower and more persistent adjustment is consistent with longer investment cycles and relatively inelastic supply in refined nickel markets. 
Further evidence on the relative contribution of each shock to price variability across horizons is provided in the next section through a forecast error variance decomposition.

\subsection{Forecast Error Variance Decomposition}

Table \ref{tab:fevd} reports the forecast error variance decomposition (FEVD) of real metal prices across structural shocks, metals, and horizons. 
To examine how the importance of each shock has evolved as the twin transition has become increasingly important, we estimate the model separately over two subsamples of equal length: Subsample 1 (12/2006--05/2015) and Subsample 2 (06/2015--12/2023).\footnote{Each entry in Table \ref{tab:fevd} represents the median contribution of each shock to the variance of the real price at a given horizon, normalized so that rows sum to one. Normally, the median across different draws would not sum to 1 because the median is non-additive. To interpret each element of the table as a percentage, we divide each entry by the sum of its row and multiply by 100.}

Consistent with the IRF analysis, AD shocks are the main driver of real metal prices, explaining more than half of total variance in most cases. 
Their dominance reflects major macroeconomic downturns in each subsample: the Global Financial Crisis in Subsample 1 and the COVID-19 pandemic in Subsample 2. 
Supply shocks account for about one-third of price variance on impact across metals and subsamples, but their contribution typically declines over longer horizons.

The FEVD also reveals two complementary patterns. 
First, for all metals and in both subsamples, the contribution of TD shocks increases with the forecast horizon. 
The same pattern is observed in the full sample case reported in the Supplement (available from the authors upon request). 
Second, for copper and nickel, the share of real price variance explained by TD shocks rises in the more recent subsample, consistent with the growing macroeconomic relevance of the twin transition. 
This effect is particularly pronounced for copper: in Subsample 2, TD shocks become the dominant driver of price variance at all horizons beyond impact, with their contribution increasing from 12.42\% on impact to 59.52\% at the 120-month horizon, compared to 25.34\% in Subsample 1 at the same horizon.

Across metals and subsamples, MD shocks consistently account for the smallest share of price variance, rarely exceeding 10\% at any horizon, with their contribution largest on impact. 
This stability across subsamples is itself informative: it implies that the redistribution of variance from Subsample 1 to Subsample 2 -- specifically, the rise of TD shocks at the expense of AD and supply shocks -- is not driven by residual demand variation captured by MD shocks. 
The quantitative importance of each shock across specific historical episodes is examined in the following subsection through historical decompositions.

\begin{table}[h!]
\caption{Forecast error variance decomposition (FEVD) of metal prices across structural shocks and subsamples.}
\label{tab:fevd}
\centering
\resizebox{\textwidth}{!}{%
\begin{tabular}{l|cccc|cccc|cccc}
\hline
& \multicolumn{4}{c|}{Cobalt}  
& \multicolumn{4}{c|}{Copper}  
& \multicolumn{4}{c}{Nickel} \\
Horizon 
& Supply & AD & MD & TD 
& Supply & AD & MD & TD 
& Supply & AD & MD & TD \\
\hline
\multicolumn{13}{c}{\textbf{Panel A: Subsample 1: 12/2006 - 05/2015}} \\
\hline
$0$   
& 34.81 & 59.27 & 3.27 & 2.64  
& 28.18 & 62.21 & 4.36 & 5.26  
& 33.35 & 50.74 & 5.30 & 10.61 \\

$3$   
& 17.72 & 69.58 & 6.88 & 5.82  
& 16.14 & 66.82 & 3.26 & 13.78  
& 20.26 & 51.52 & 5.03 & 23.19 \\

$12$  
& 10.27 & 70.52 & 7.67 & 11.53  
& 15.62 & 55.74 & 4.08 & 24.56  
& 12.90 & 51.84 & 5.51 & 29.75 \\

$24$  
& 9.19 & 56.81 & 6.58 & 27.43  
& 16.79 & 54.12 & 4.22 & 24.87  
& 12.67 & 49.66 & 5.50 & 32.17 \\

$120$ 
& 9.14 & 52.07 & 5.84 & 32.95  
& 16.98 & 53.43 & 4.26 & 25.34  
& 12.59 & 48.87 & 5.44 & 33.10 \\

\hline
\multicolumn{13}{c}{\textbf{Panel B: Subsample 2: 06/2015 - 12/2023}} \\
\hline
$0$   
& 22.08 & 58.14 & 12.98 & 6.80  
& 29.54 & 47.15 & 10.90 & 12.42  
& 29.62 & 55.48 & 9.22 & 5.68 \\

$3$   
& 29.79 & 58.84 & 6.42 & 4.96  
& 23.26 & 32.77 & 6.91 & 37.06  
& 23.21 & 62.44 & 6.83 & 7.52 \\

$12$  
& 25.82 & 57.64 & 5.48 & 11.07  
& 17.17 & 18.11 & 7.34 & 57.37  
& 15.33 & 52.26 & 7.34 & 25.06 \\

$24$  
& 26.36 & 53.45 & 5.31 & 14.89  
& 15.58 & 18.25 & 7.22 & 58.96  
& 14.02 & 45.51 & 7.72 & 32.75 \\

$120$ 
& 25.83 & 52.13 & 5.30 & 16.74  
& 15.48 & 17.82 & 7.18 & 59.52  
& 13.27 & 43.39 & 7.66 & 35.68 \\

\hline
\end{tabular}}
\caption*{\scriptsize \textit{Notes}: Each number represents the median percentage contribution of each shock to the variance of the real price for the corresponding horizon. Since the median is non-additive, taking the median across draws would result in rows that do not necessarily sum to one. For this reason, the numbers have been normalized by dividing each entry by the sum of its row and multiplying by 100.}
\end{table}

\subsection{Historical Decompositions}\label{sec:histdecomp}
We rely on  historical decompositions to attribute observed price movements to specific structural shocks over time. 
We focus on two price surge episodes shared across all three metal markets: the post--Global Financial Crisis period (January--June 2009) and the post-pandemic surge (December 2021--June 2022), which marked a turning point driven by rapid electric vehicle adoption and the accelerated deployment of renewable energy technologies \citep[see e.g.][]{IEAbattery,iea_ev_sales_2025}.\footnote{According to \citet{iea_ev_sales_2025}, total sales of EVs (including both plug-in hybrid electric vehicles and battery electric vehicles) more than tripled between 2020 and 2022, from around 3 million to 10.1 million units.}

The left panels of Figure \ref{fig:HD} show that the 2009 recovery was largely driven by AD shocks (light blue bar) across all three metals after March 2009, with supply shocks (blue bar) contributing to upward pressure for copper and nickel. 
TD shocks (orange bar) are generally small and in some cases negative; TD shares of total explained variation are 7.8\% for copper and 9.3\% for nickel.\footnote{To quantify the relative role of TD shocks in each episode, we compute their share of total explained price variation as $\sum_{t \in \mathcal{T}} |\hat{\psi}_{\mathrm{TD},t}| \,/\, \sum_{j} \sum_{t \in \mathcal{T}} |\hat{\psi}_{j,t}|$, where $\hat{\psi}_{j,t}$ is the contribution of shock $j$ at time $t$, $\mathcal{T}$ denotes the episode window, and absolute values prevent positive and negative contributions from canceling.}

\begin{figure}[h!]
    \centering
    \caption{Historical decomposition of the real prices of cobalt, copper, and nickel.}
    \includegraphics[width=0.8\linewidth]{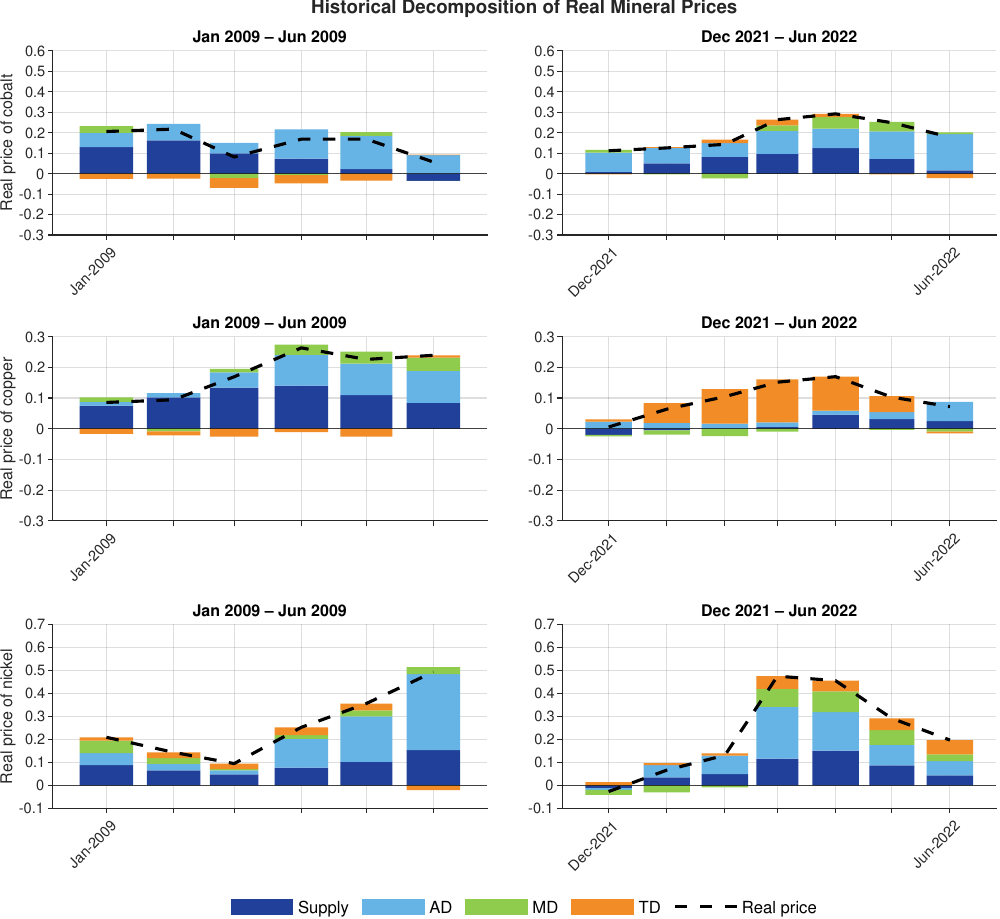}
   
    \label{fig:HD}
\end{figure}

Turning to the right panels, AD shocks remain the primary driver of the post-pandemic price surge for cobalt and nickel, while supply shocks also contribute positively after March 2022.
For nickel, MD shocks (green bar) reflect heightened precautionary demand linked to Russia’s invasion of Ukraine, which triggered fears of supply losses from a key producer of high-grade nickel. 
Consistent with the definition of MD shocks, these concerns, rather than realized shortages, contributed to the surge in real nickel prices from March 2022 onwards, as market participants revised expectations and engaged in anticipatory stockpiling \citep{ReutersNickel}.

Finally, TD shocks (orange bar) in the most recent subsample play a quantitatively significant important role for copper and, to a lesser extent, nickel, consistent with rapid EV adoption and the accelerated deployment of renewable energy technologies \citep[see e.g.][]{IEAbattery,iea_ev_sales_2025}. 
The TD share for nickel rises from 9.3\% to 14.2\%, while for copper the increase is more pronounced, from 7.8\% to 57\%, making TD shocks the most important driver of copper price movements in the post-pandemic episode.

\section{Conclusions}\label{sec:concl}
In this paper, we construct a novel monthly metal-specific derived demand indexes for cobalt, copper, and nickel using Google Trends web search data, and incorporate them into metal-specific SVAR models to identify supply- and demand-side shocks in global metal markets. 
Our key contribution is the identification of transition demand (TD) shocks associated with the demand for technologies crucial for the energy and digital transitions.

The SVAR model analysis shows that TD shocks generate persistent price effects that build gradually and decay slowly, especially for copper and nickel, whereas supply and metal-specific demand shocks tend to have more immediate and less persistent effects.
Forecast error variance decompositions and historical decompositions analyses show that TD shocks have become more relevant in recent years, especially for copper and nickel, highlighting the growing macroeconomic relevance of transition-related demand. 


Emerging sources of demand, including artificial intelligence and data centre infrastructure, may further strengthen the role of metals such as copper. 
As these forces are only partially captured in our indexes and have become more relevant in recent years, our estimates may understate the broader contribution of transition-related demand.

Overall, not all demand-driven price increases are alike: distinguishing among these shocks is essential for understanding commodity price dynamics in an economy undergoing rapid energy and digital transitions.

\pagebreak

\bibliography{biblio}

\pagebreak

\appendix
\setcounter{table}{0}
\renewcommand{\thetable}{A\arabic{table}}
\setcounter{figure}{0}
\renewcommand{\thefigure}{A\arabic{figure}}

\end{document}